\documentclass[preprint2]{emulateapj}
\usepackage{amsmath,amstext,gensymb,color}
\usepackage[]{subfigure}
\usepackage{enumitem}
\usepackage[]{hyperref}
\hypersetup{colorlinks=false,
		linkcolor=black,
		urlcolor=cyan}
        
\def\kms{km\,s$^{-1}$}
\def\Ha{H{$\alpha$}}
\def\Hb{H{$\beta$}}

\def\M{M$_{\odot}$}

\def\ae{SN~2015bn}

\shorttitle{SN~2015bn}
\shortauthors{Inserra et al.}

\begin{document}

\title{Spectropolarimetry of superluminous supernovae: insight into their geometry}
\author{
C. Inserra\altaffilmark{1},
M. Bulla\altaffilmark{1},
S. A. Sim\altaffilmark{1},
and S. J. Smartt\altaffilmark{1}.}

\altaffiltext{1}{Astrophysics Research Centre, School of Mathematics and Physics, Queens University
  Belfast, Belfast BT7 1NN, UK; c.inserra@qub.ac.uk}
  
\begin{abstract}
We present the first spectropolarimetric observations of a hydrogen-free superluminous supernova at \mbox{$z=0.1136$}, namely \ae. The transient shows significant polarization at both the observed epochs: one 24 days before maximum light in the rest-frame, and the subsequent at 27 days after peak luminosity. Analysis of the Q-U plane suggests the presence of a dominant axis and no physical departure from the main axis at either epoch. 
The polarization spectrum along the dominant axis is characterized by a strong wavelength dependence and an increase in the signal from the first to the second epoch. We use a Monte Carlo code to demonstrate that these properties are consistent with a simple toy model that adopts an axi-symmetric ellipsoidal configuration for the ejecta. We find that the wavelength dependence of the polarisation is possibly due to a strong wavelength dependence in the line opacity, while the higher level of polarisation at the second epoch is a consequence of the increase in the asphericity of the inner layers of the ejecta or the fact that the photosphere recedes into less spherical layers. The geometry of the superluminous supernova results similar to those of stripped-envelope core collapse SNe connected with GRB, while the overall evolution of the ejecta shape could be consistent with a central engine.  

\end{abstract}

\maketitle

\section{Introduction}\label{sec:intro}

The last six years have seen the surprising discovery of new classes of intrinsically bright supernovae. They show absolute peak magnitudes of M$\sim-21$ \citep[e.g.][]{qu11,2011ApJ...743..114C,in14} and a tendency to occur in dwarf, metal poor galaxies \citep[e.g.][]{ch13,lu14,le15a,ch16}. They do not exhibit the typical narrow spectroscopic features of strongly interacting supernovae and they are usually referred as superluminous supernovae \citep[SLSNe; see the review of][]{gy12}.  

They can be divided in two main groups, hydrogen free - and hence labeled SLSNe I - and hydrogen rich - therefore called SLSNe II. The first class is better studied: they show blue continua at maximum light, a distinctive ``W'' feature due to O~{\sc ii} at early epochs and, at about 30 days after peak they are spectroscopically similar to normal or broad-lined type Ic SNe at peak luminosity \citep{pa10}; consequently, they are also usually labeled SLSNe Ic \citep[e.g.][]{in13}.
Additionally, SLSNe Ic show different light curve behavior and, as a consequence, have been divided in the subgroup of fast evolving \citep[e.g. SN~2005ap, SN~2010gx;][]{pa10,qu11} and that of slow evolving \citep[e.g. SN~2007bi, PTF12dam;][]{gy09,ni13}. The SLSNe of type II are fewer in number and they also show blue continua at maximum light, together with broad hydrogen features. At about 20 days after peak they show some resemblances to normal type IIL, although by definition they are several magnitudes brighter  \citep[see][for a review]{in16}.

Despite the increasing number of objects found every year, the nature of their explosion and the progenitor scenario are still debated. The favored scenario for all the types is that of an explosion driven by a magnetar as a central engine, which deposits energy into the supernova ejecta and significantly enhances the luminosity \citep{wo10,kb10,de12}. However, alternative scenarios such as the accretion onto a central black hole \citep{dk13}, the interaction with a dense circumstellar medium \citep[CSM,][]{ch12} and a pair instability explosion  are still feasible alternatives.
The pair-instability mechanism is only physically plausible for some of the objects, in particular those with the slowest 
evolving lightcurve  \citep[e.g. see][]{gy09,koz15} 

A powerful diagnostic to distinguish between scenarios 
is 
polarimetry since it can unveil information on the geometry of the explosion and hence increase our understanding of the transients.
Imaging polarimetry of a fast evolving SLSNe Ic has been reported by \citet{le15} but no evidence of asymmetries was found. However, spectropolarimetry offers a more in depth analysis of the geometry of SN explosions \citep[see][for a review]{ww08}. 
Quantitative modelling of the lightcurves of  SLSNe Ic indicates that the data are well 
matched by the explosion of massive progenitors with a central engine 
\citep[e.g. as in ][]{in13,ni13}. If this were true, we would expect to see an intrinsically asymmetric explosion characterized by a dominant polarization angle as observed for stripped envelope SNe \citep[e.g.][]{wa01,ma07,tan12}. Indeed many
core collapse SNe  show large-scale departures from axisymmetry \citep[e.g.][]{le06,ww08}. 

If a magnetar is the central engine that powers the extreme luminosity,, then 
 the strong magnetic field (B$\sim10^{14}$G) and rapid rotation could lead to asymmetries \citep{chk16}. 
Such asymmetries are potentially stronger than in normal stripped envelope SNe. Detection of such signatures
may suggest a core-collapse origin  combined with  an asymmetric magnetar energy injection process. On the other hand, alternative scenarios such as the interaction with a CSM would exhibit spectropolarimetric evolution similar to those of type IIn SNe \citep[e.g.][]{le00,wa01}.  These signatures are different than that of stripped envelope SNe, since they typically show
unpolarised broad lines and loops across spectral lines arising from the CSM \citep{ww08}. 

An axisymmetric ejecta could be the consequence
of: an aspherical production of energy and momentum from an explosion due to magnetohydrodynamic jets \citep{ko99} or magnetoturbulence \citep[e.g.][]{mo15}, accretion flow around the central neutron star \citep[e.g.][]{chev89}, asymmetric neutrino emission \citep[e.g.][]{wha10,mu15} or a combination of these; or the fact that the material could be ejected in clumps.

In this paper we present our spectropolarimetric observations, as well as the interpretation of the geometry of the brightest
known superluminous supernova, namely \ae\/, which belongs to the group of slow evolving SLSNe Ic.

\section{Superluminous supernova \ae}
\ae\/ was discovered by the Catalina Real-time Transient Survey \citep[CRTS,][with IDs CSS141223-113342+004332 and MLS150211-113342+004333]{dr09} on  23  December 2014 and subsequently by the Pan-STARRS Survey for Transients \citep[PSST, with ID PS15ae,][]{hub15} on 2015 February 15 UT. It was classified on the 17 February 2015 by the the Public ESO Spectroscopic Survey for Transient Objects \citep[PESSTO,][]{2015A&A...579A..40S} as a type Ic superluminous supernova at $z=0.11$ similar to several SLSNe Ic around maximum light \citep{at7102}. The main PESSTO follow-up campaign is presented by \citet{ni16}, where the redshift $z=0.1136$ has been confirmed from the detection of narrow lines from the host galaxy. Adopting  H$_0=72$ \kms\/,  $\Omega_{\rm M}=0.27$ and $\Omega_{\Lambda}=0.73$ as standard cosmological parameters, this corresponds to $d_{\rm L}=514.3$ Mpc. \citet{at8552} reported that no radio continuum emission from the SN was detected by the VLA on 2015 December 11.67 UT, with a 3$\sigma$ upper limit of 40 $\mu$Jy at a mean frequency of 21.85 GHz. This corresponds to a luminosity limit of of L$_{\nu}\lesssim 1.3\times 10^{28}$ erg s$^{-1}$ Hz$^{-1}$ at $z=0.1136$, which sets a limit on the mass-loss rate (${\rm \dot{M}<10^{-1.8}}$ \M\/ yr$^{-1}$) of the CSM in the case of the interaction scenario \citep{ni16}.

\begin{deluxetable*}{lcccccc}
\tablewidth{0pt}
\tablecaption{ \label{table:log}}
\tablehead{
\colhead{Date}& \colhead{MJD}& \colhead{Phase\tablenotemark{a}} & \colhead{g\tablenotemark{b}} & \colhead{Exposure Time}& \colhead{S/N\tablenotemark{c}}& \colhead{Mean Airmass} \\
\colhead{} & \colhead{} & \colhead{(days)} & \colhead{(mag)} &  \colhead{(s)}}
\startdata
2015/02/22 & 57076.23 & -23.7 & 16.82 & 12 x 800 & 460 &1.1\\
2015/04/20 &  57132.99 & +27.5 & 17.34 & 12 x 800 & 390 &1.3
\enddata
\tablecomments{\tablenotetext{a}{phase with respect to the g-band maximum, corrected for time dilation}
\tablenotetext{b}{$g$ magnitudes are derived from {\sc sms} \citep{in16}.}
\tablenotetext{b}{S/N in the central wavelength region, bluer wavelengths have higher S/N, while redder smaller.}}
\end{deluxetable*}

\section{Observation and data reduction}\label{sec:data}
Spectropolarimetric observations of \ae\/ were conducted with the Very Large Telescope (VLT) + the FOcal Reducer/low dispersion Spectrograph 2 (FORS2)\footnote{Mounted on the UT1 (Antu) Cassegrain focus.}  on the 22 February 2015 and 20 April 2015, which correspond to $-23.7$d and $+27.5$d with respect to the maximum light in the $g$-band \citep[MJD 57102.5,][]{ni16}. Both epochs are close to the broad peak of the SLSN light-curve.  A log of the observations is given in Table~\ref{table:log}. We used a half-wave retarder plate (HWP) at four angles (0, 22.5, 45 and 67.5 degrees) through three iterations per epoch. All spectra were observed with the grism 300V and blocking filter GG453 in order to avoid second order contamination, giving an observed spectral range of 4400 -- 9200~\AA,  and a dispersion of 1.68~\AA\,pix$^{-1}$. A slit width of 1.0 arcsecs resulted in a resolution of 13~\AA.
Data were reduced in a standard fashion (including trimming, overscan, bias
correction and flat-fielding) using standard routines within
IRAF\footnote{Image Reduction and Analysis Facility, distributed by
  the National Optical Astronomy Observatories, which are operated by
  the Association of Universities for Research in Astronomy, Inc,
  under contract to the National Science Foundation.}. 

Optimal extraction of the spectra was adopted. In addition, ordinary and extraordinary beams were processed separately. Wavelength calibration was performed
using spectra of comparison lamps acquired with the same
configurations as the SN observations. Atmospheric extinction
correction was based on tabulated extinction coefficients for the telescope site.
Flux calibration was performed using
spectro-photometric standard stars observed on the same nights with
the same set-up as the SNe. The flux calibration was also checked by comparison with the photometry provided in \citet{ni16}, integrating the spectral flux transmitted by standard {\it griz} filters  - using the {\sc python} programme {\sc sms} \citep{in16} - and adjusted by a multiplicative factor when necessary. The resulting
flux calibration matches the measured photometry to within 0.1 mag. Instrumental polarization and the position angle offset were checked by observing polarized and unpolarized standard stars during the second observation.

\begin{figure*}
\includegraphics[width=18cm]{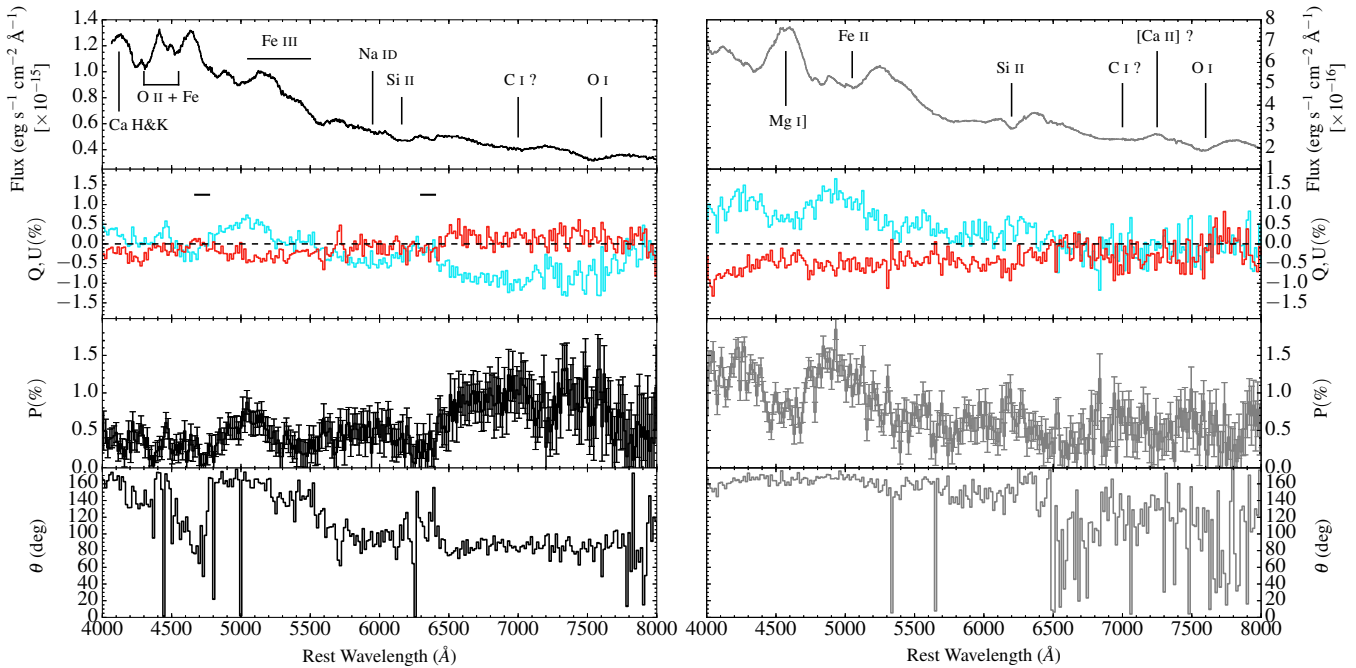}
\caption{From top to bottom are displayed the flux spectrum, the Stokes parameters Q (light blue) and U (red), the polarization vector with errors and the polarization position angle (not corrected for ISP). On the left the data of the first (-23.7d) epoch, while on the right are displayed those of the second (+27.5d) epoch.
The horizontal black lines refer to the 100~\AA\/ regions where we calculated the Stokes parameters for the ISP.} 
\label{fig:original}
\end{figure*}

\section{Interstellar polarization and errors}\label{sec:isp}

Stokes parameters were computed ($Q$ and $U$)  and error estimates were made
using the procedures described by \citet{pr06}. The Stokes parameters were re-binned to 15~\AA\/ to increase the signal to noise (S/N). 

The Galactic reddening toward the position of the SN is E(B-V)~=~0.02 \citep{sf11}. No Na~{\sc id} $\lambda\lambda$5890, 5896 from the host galaxy was detected in our spectra, which is consistent with SLSNe \citep[e.g.][]{ch13,in14,vr14,in16} since they usually do not have reddening from the dwarf host galaxies. We will assume a total reddening equal only to the Galactic component. The assumption of both low host galaxy and Galactic reddening is supported by (i) the color of the spectra - T$_{\rm BB}\sim$10000 K and $\sim$9400 K for the first and second epoch, respectively - which are similar to those of comparable SLSNe Ic at around the broad peak phase \citep{in13,ni16}, and (ii) 
the measurement of the host galaxy narrow emission lines of \Ha, \Hb\/ and the resulting Balmer decrement, which points to E$_{\rm Host}$(B-V)~=~0 \citep{ni16}.

For the adopted reddening, the Serkowski-Whittet law allows us to place a limit of P$_{\rm ISP}< 0.2\%$ \citep[][]{ser75,wh92} 
on the interstellar polarization (ISP) due to intervening dust in the line of sight to 
\ae\/.
In addition, we found three stars from the \citet{hei00} catalog within 3 degrees of the SN position, which have P~$=0.07\%$, P~$=0.10\%$ and P~$=0.05\%$ corresponding to a mean of $\bar{\rm P}=0.07\pm0.04\%$, in full agreement with the low Galactic ISP along the line of view.

Furthermore, as an independent test, we have derived the ISP value directly from our spectropolarimetric observations assuming that the intrinsic polarization in regions of strong emission lines or line blanketing is negligible \citep{tr93}. With this assumption, polarization signals in these spectral regions can be attributed to the ISP. We identified the region of the emission of Si~{\sc ii} $\lambda$6355 and the region around 4500~\AA\/ affected by Fe~{\sc ii} and Fe~{\sc iii} blending (see Section~\ref{sec:sp}) as suitable for such analysis. The Stokes parameters were calculated over a range of 100~\AA\/ in the binned (15~\AA) regions of 4670-4770 \AA\/ and 6300-6400 \AA\/. In the first epoch we measured Q$_{\rm ISP}=-0.15\%$ and U$_{\rm ISP}=-0.06\%$, which gives us a polarization level P$_{\rm ISP}=0.16\%$ consistent with the previous limits. For our analysis, we applied this correction to the spectra of both the first and the second epoch -- as the latter does not present the iron blanketing.

To evaluate the errors on our polarimetry, we measured the root mean square (rms) of the normalized flux differences of our ordinary and extraordinary beams \citep[see eq. 4 and 5 in][]{pr06} and then propagate those to the Stokes parameters and the polarization vector.
Since in our data set ${\rm \eta=P_{0}(S/N)} \sim 8$, where P$_{0}$ is the input polarization and S/N is the signal to noise ratio \citep{sa99,pr06}, we can check the validity of our error estimates by comparing them to the analytic expression for the absolute error in P given by \citet[][see their eq. 10]{pr06}.
We find that the analytic error estimates are in good agreement with our error calculations, which we adopt throughout the following. Furthermore, due to the large $\eta$ value, we can safely assume that both P and $\sigma_{\rm P}$ follow a Gaussian distribution and do not require debiasing corrections \citep{sa99}.

\section{Spectroscopic properties}\label{sec:sp}

The spectroscopic evolution of SLSNe Ic has been well sampled \citep[e.g.][]{pa10,in13}, especially that of slow-fading ones \citep[e.g.][]{ni13,vr14,ni16} with the evolution from pre- to post-peak spectra mainly showing the disappearance of the characteristic O~{\sc ii} lines soon after peak and the decrease of ionization state of the metal lines. As already shown in \citep{ni16}, \ae\/ follows such evolution. 

Our first spectrum 
was taken
24 days before peak magnitude in g band. A black-body fit to the available spectral range yields T$\sim$10000 K, which corresponds to a radius of $6.5\times10^{15}$cm and black-body peak bluer (2897.8~\AA) than covered by our wavelength range. The spectrum is dominated by metal lines: the Ca~H\&K lines are visible on the blue edge of our spectrum,
redward of which are two double peak absorption features at 4400~\AA\/ and 5000~\AA, most likely due to a combination of Fe~{\sc iii} and Fe~{\sc ii} with a possible contribution of O~{\sc ii}, which could be excited by non-thermal processes \citep{maz16}. Redward, the spectrum is again dominated by iron up to 5500~\AA\/, while Na~{\sc id} and Si~{\sc ii} $\lambda$6355 are visible in the center part around 6000~\AA.
The region around 7500~\AA\/ shows a shallow O~{\sc i} absorption line, together with another shallow P-Cygni profile at 7200~\AA\/ present only in the slow-fading SLSNe Ic and tentatively identified as C~{\sc i} \citep[see fig.~11 in][]{ni16}. Following the modeling of SLSNe Ic spectra reported by \citet{maz16} an alternative identification for the 7200~\AA\/ line could be O~{\sc ii}.

The second spectrum - taken 28 days after the maximum light - has a slightly lower black-body temperature (T$\sim$9400 K) and larger black-body radius $8\times10^{15}$cm compared to the first. At this epoch Fe~{\sc iii} has been replaced by Fe~{\sc ii} (see top panels of Fig.~\ref{fig:original}). The noticeable changes are in the region blueward of Si~{\sc ii}. Mg~{\sc i}] $\lambda$4571 appeared in emission and replaced Fe~{\sc iii} and - together with the increase of Fe~{\sc ii} features with respect to the previous epoch - has reshaped the profile of that region that no longer shows the two absorption features. 
The other distinguishable change is the shift to the red of the broad profile around 5300~\AA\/ that is due again to Fe~{\sc ii} lines replacing their higher ionized ions. The other two P-Cygni profiles redder than Si~{\sc ii} are still present and due to the same elements of the first epochs, even if the feature around 7200~\AA\/ could now have a small contribution from [Ca~{\sc ii}] $\lambda\lambda$7291,7323 as also shown by \citet{ni16}.

\section{Spectropolarimetry}\label{sec:spol}

Our spectropolarimetric observations of \ae\/, not corrected for the ISP, are shown in Fig.~\ref{fig:original}.
In both our spectra we detect clear polarization signals. In the first epoch, the typical degree of polarisation increases from $\sim0.5$\% in the blue to a maximum of P$\sim1.2$\% at $\sim7000 - 7500$~\AA. However, in the post-maximum spectrum we find the opposite behavior: a decrease of the polarization from the blue (P$\sim1.4$\%) to the red (P$\sim0.5$\%). 

Most of the peaks in the polarization signal are roughly coincident with the absorption minima of identifiable P-Cygni profiles in the spectra: at the first epoch, there is a clear increase in polarization across the Fe~{\sc iii} absorption feature at about 5100~\AA\/ and also across both the oxygen lines at 
4400~\AA\/ and 7700~\AA\/ . At the second, post-maximum epoch (+27.5d), the degree of polarization across O~{\sc i} (P$\sim0.9$\%) remains similar to that found at the first epoch (P$\sim$1.2\%). However, there are significant changes to all the features blueward of O~{\sc i}. The Si~{\sc ii} line, which shows a more prominent P-Cygni profile in the +27.5d flux spectrum than the -23.7d, reaches a peak polarization signal of $\sim$1\%, while the Fe~{\sc ii} 5000~\AA\/ features reaches P$\sim1.7$\%.

The high-polarization feature at about 7000~\AA\/ corresponds to the blue end of a shallow absorption trough in both the spectra, which was tentatively identified as carbon by \citet{ni16} or it could alternatively be associated with oxygen (see Section~\ref{sec:sp}).

The general evolution from pre-maximum to post maximum phase shows an increase of polarization - especially for the metal lines at blue wavelengths - as usually observed in stripped envelope SNe \citep[e.g.][]{wa03,ma07,re16}.

\subsection{Dominant axis and asymmetry}\label{sec:rot}
\begin{figure}
\includegraphics[width=\columnwidth]{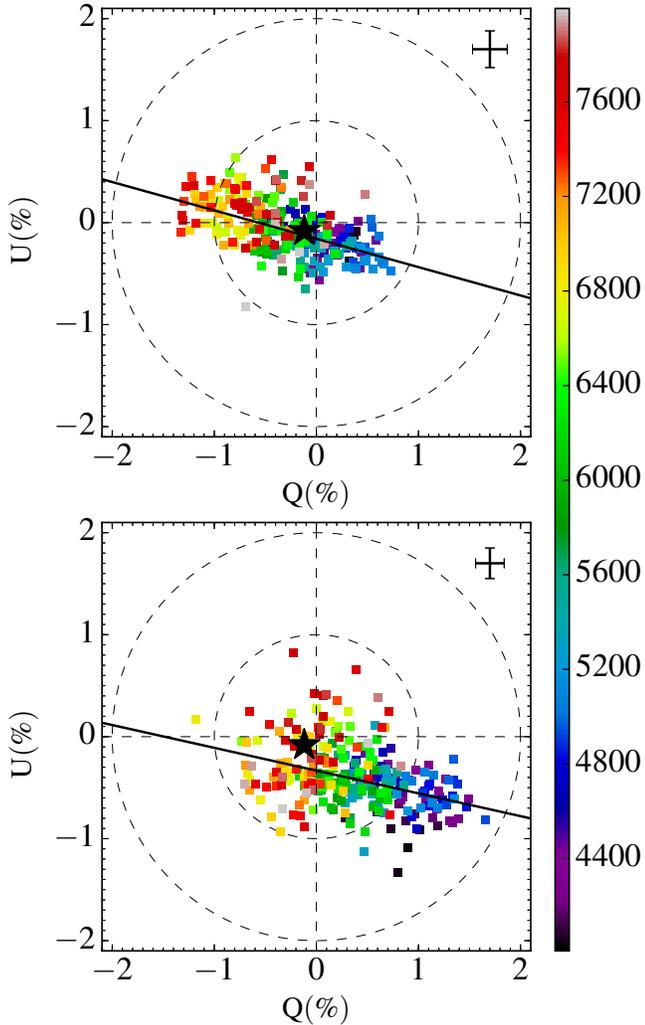}
\caption{QU plane of \ae\/ at -23.7 days from peak (top) and +27.5 (bottom). Wavelengths are shown from blue to red colors. Concentric dashed circles are equivalent to P~$=1$\% and P~$=2$\% from the inner to the outer. The black stars represents the ISP in the QU plane (see Section~\ref{sec:isp}). The black solid line shows the best linear fit representing the dominant axis. Average errors for the data are indicated by the cross bars shown in the top corner of each panel.} 
\label{fig:QU}
\end{figure}

In Fig.~\ref{fig:QU} we show our
spectropolarimetric dataset in the Q-U plane. 
As before, we note that the polarization reaches a maximum of about 1.4\% in the first epoch and almost 2\% in the second.

\begin{figure}
\includegraphics[width=\columnwidth]{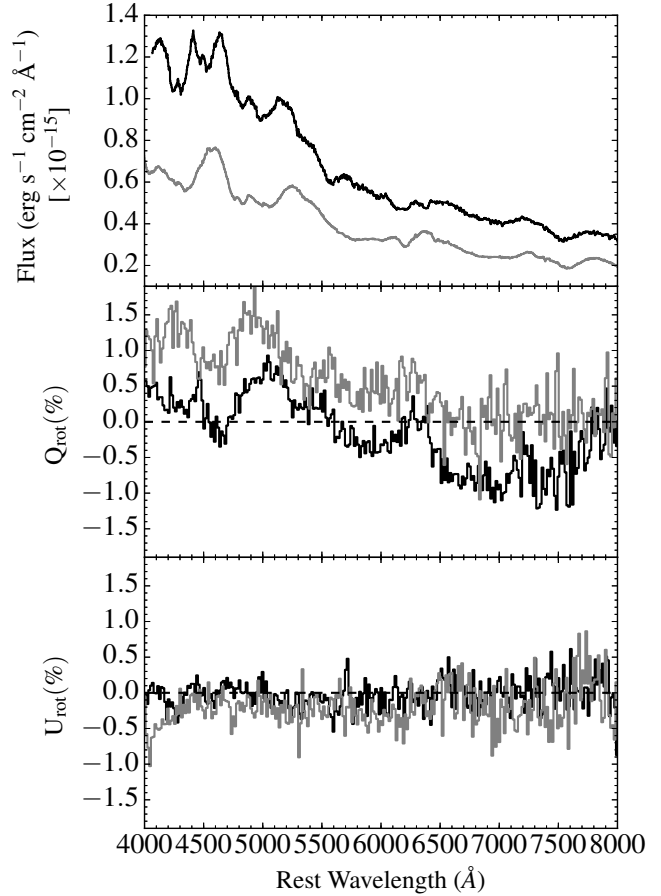}
\caption{Top: flux spectra for the first (black) and second (gray) epoch. Middle: the polarization along the dominant axis. Bottom: the polarization along the orthogonal axis.} 
\label{fig:rot}
\end{figure}

The distribution of our data points in the Q-U plane clearly suggests a preferred direction, which we identify by fitting the data with a least squares fit weighted by the observational errors
following the procedure of \citet{wa03}. The best fits to the available dataset give a dominant axis with a position angle ($\theta_{\rm d}=1/2\times$arctan$\beta$) of -8$^{\degree}\pm1$ for the pre-maximum and -6$^{\degree}\pm1$ for the post maximum epoch with respect to a North-South axis on the night sky\footnote{The 0$^{\degree}$ angle means that the axis is aligned with the N/S direction. A negative angle means that the dominant axis is rotated from North to West.}.
We note that, in the first epoch, the dominant axis almost cross the origin and passes through the ISP measurement (black star in Fig.~\ref{fig:QU}), suggesting that our estimate for the low level of ISP was correct. On the other hand, the best fit of the second epoch is slightly different than the previous and does not cross the ISP. In principle, a difference might indicate an 
intrinsic change of the dominant axis from the first to the second epoch and hence a different dominant orientation between the external and inner layers of the ejecta. 
Alternatively, the difference between the fits can be due to the larger uncertainties on Q and U in the second epoch, especially in the red part of the spectrum. 
However, the change in angle of the best fit between the two epoch is sufficiently small, and comparable with the errors, that it can be attributed to measurement uncertainties only. We note that if we force our fit to pass through the ISP in the second epoch we obtain an angle of  -11$^{\degree}\pm1$, which is still comparable to that of the first epoch.

We note that our analysis clearly suggests that the data at the first epoch cross the origin in the Q-U plane (from second to fourth quadrant). As discussed by \cite{pat10}, this often suggests that the ISP may have been inaccurately estimated and that the true value of the ISP should be at one of the two extremes of the data distribution in the Q-U plane.
In our case, however, that would imply a 1\%$<$P$_{\rm ISP}$$<$2\%, which is inconsistent with the low reddening observed toward the SN (see Sect.~\ref{sec:isp}). Thus we favor an interpretation in which 
the data distribution in the Q-U plane, including the crossing between quadrants, is intrinsic to the ejecta and not a consequence of significantly underestimated ISP.

It is convenient to project the polarization vectors in a new coordinate system that has a component along the dominant axis
\begin{equation}
{\rm Q}_{\rm rot} = ({\rm Q} - {\rm Q}_{\rm ISP}) \cos(2\theta) + ({\rm U} - {\rm U}_{\rm ISP}) \sin(2\theta) ,
\end{equation}
which represents global geometric deviations from spherical symmetry, and an orthogonal component to the dominant axis 
\begin{equation}
{\rm U}_{\rm rot} = - ({\rm Q} - {\rm Q}_{\rm ISP}) \sin(2\theta) + ({\rm U} - {\rm U}_{\rm ISP}) \cos(2\theta) ,
\end{equation}
which shows physical deviations from the dominant axis \citep[for further details see][]{ww08, le01}. In Fig.~\ref{fig:rot} we show our polarization vectors in the new system. For both epochs we used a single angle of $\theta=8^{\degree}$.
This makes clear that, while our data show a strong deviation from spherical symmetry (seen in Q$_{\rm rot}$), there is only weak evidence of physical deviation around the dominant axis (i.e. U$_{\rm rot}$ is fairly consistent with zero, suggesting that the ejecta are approximately axisymmetric).
At both epochs, Q$_{\rm rot}$  
shows a wavelength dependence decreasing from blue to the red. At the first epoch, we find a clear sign reversal with Q$_{\rm rot}$ changing from positive values for blue wavelengths ($\lesssim 6000$~\AA) to negative values in the red. The second epoch is more polarized than the first and does not show a sign reversal.

\ae\/ would be classified as spectropolarimetry type D0 in the terminology of \citet{ww08} since it shows a distinct dominant axis and a distribution orthogonal to that consistent with observational noise. In addition, the stronger lines do not exhibit a loop in the Q-U diagram, hence they do not show large significant changes in amplitude and position angle.

\subsection{Comparison with LSQ14mo}

To date, polarimetric data has been presented for only one other SLSNe Ic, namely LSQ14mo \citep{le15}. However, only broad-band polarimetry in the FORS2 V filter was obtained when the object had an observed magnitude $20.94< g<19.38$. \citet{le15} reported an average polarization of P~=~$0.52\pm0.15$\% over five epochs from -7d to 18d since maximum light, which could be explained by ISP in the host galaxy. A similar result, based again on broad-band imaging polarimetry, has been reported for a single epoch of SLSN II PS15br by \citet{in16}. 

We checked if the findings of \citet{le15} are consistent with ours.
To do that, we scaled \ae\/ spectra to the same distance of LSQ14mo ($d_{\rm L}=1266.4$ Mpc). This decreased our observed flux by a factor $\sim7$ - corresponding to a luminosity $\sim2$ mag fainter with respect to those listed in Tab.~\ref{table:log} - and the S/N by a factor of almost three.
Furthermore, our total exposure time is almost double than that used for LSQ14mo, which would imply an additional decrease of the S/N by a factor of a third and hence higher error and scatter to the previous results.

Integrating over the VLT+FORS2 V broad-band we evaluated P$_1$ = $0.44\pm0.94$\% and P$_2$ = $0.91\pm0.79$\% for the first and second epochs, respectively. These values have relative errors of 200\% and 87\% and are similar to those reported by \citet{le15} for their first (-6.9d) and last (+18.5d) epochs. The average polarization is P~=~$0.67\pm0.86$\%. Hence, the data quality in case of LSQ14mo was likely to have been too low to detect the intrinsic polarimetric signal of the SN, preventing any conclusion about the geometry. 

In addition, \ae\/ was significantly brighter than LSQ14mo
( $M_{\rm g}^{\rm peak}=-22.0$ for \ae\/ and $M_{\rm g}^{\rm peak}=-21.2$ for LSQ14mo). 
Therefore the broad-band observations of  LSQ14mo (which have less 
than 3 hr of integration) almost certainly have 
insufficient signal-to-noise to determine reliable Stokes parameters and the polarization vector. 
Therefore it is not surprising that \citet{le15} could not draw any firm conclusion about the intrinsic polarization of LSQ14mo from their observations. 

Furthermore, with imaging polarimetry any information about lines and continuum contribution from the ejecta is omitted and hence the results are hard to interpret. Indeed, during the time baseline available for \ae\/ and LSQ14mo, the region of interest (3900-5500~\AA) has experienced a significant spectral change:  increase in strength of iron lines, as well as the appearance of Mg~{\sc i]} $\lambda$4571 in the latest epoch and the disappearance of O~{\sc ii} lines from the first.

\begin{deluxetable*}{cccccccccccccccccc}
\tablewidth{0pt}
\tablecaption{ \label{table:tardis}}
\tablehead{
\colhead{Model}& \colhead{Epoch}& \colhead{Log$_{10}$\,L$_\textrm{bol}$} & \colhead{v$_\textrm{ph}$} & \colhead{v$_\textrm{out}$} & \colhead{$\rho_\textrm{ph}$} & \colhead{He}& \colhead{C} & \colhead{O} & \colhead{Ne}& \colhead{Mg}& \colhead{Si} & \colhead{S} & \colhead{Ca}& \colhead{Ti} & \colhead{Fe} &  \colhead{Co} & \colhead{Ni}\\
& & \colhead{erg s$^{-1}$} & \colhead{km s$^{-1}$} & \colhead{km s$^{-1}$} & \colhead{g cm$^{-3}$} & & & & & & & & & & & &
}
\startdata
A & 02/22 & 44.2 & 8e8 & 1.6e9 & 7.3e-14 & 0.10& 0.40& 0.475& 0.02&7e-4 & 0.002& 5e-4 &5e-5 &2e-5 &5e-4 &3e-4 &1e-5 \\
A & 04/20 & 44.1 & 8e8 & 1.6e9 & 3.6e-14 & 0.10& 0.40& 0.475& 0.02&5e-4 & 0.002& 5e-4 &5e-5 &2e-5 &5e-4 &3e-4 &1e-5 \\

B & 02/22 & 44.2 & 8e8 & 1.6e9 & 7.3e-14 & 0 & 0 & 0 & 0 & 0 & 0 & 0 & 0 & 0 &  0.414 & 0.585 & 3.5e-4 \\
B & 04/20 & 44.1 & 8e8 & 1.6e9 & 3.6e-14 & 0 & 0 & 0 & 0 & 0 & 0 & 0 & 0 & 0 & 0.626 & 0.374 & 1e-6
\enddata
\end{deluxetable*}

\begin{figure*}
\begin{center}
\includegraphics[width=0.87\textwidth]{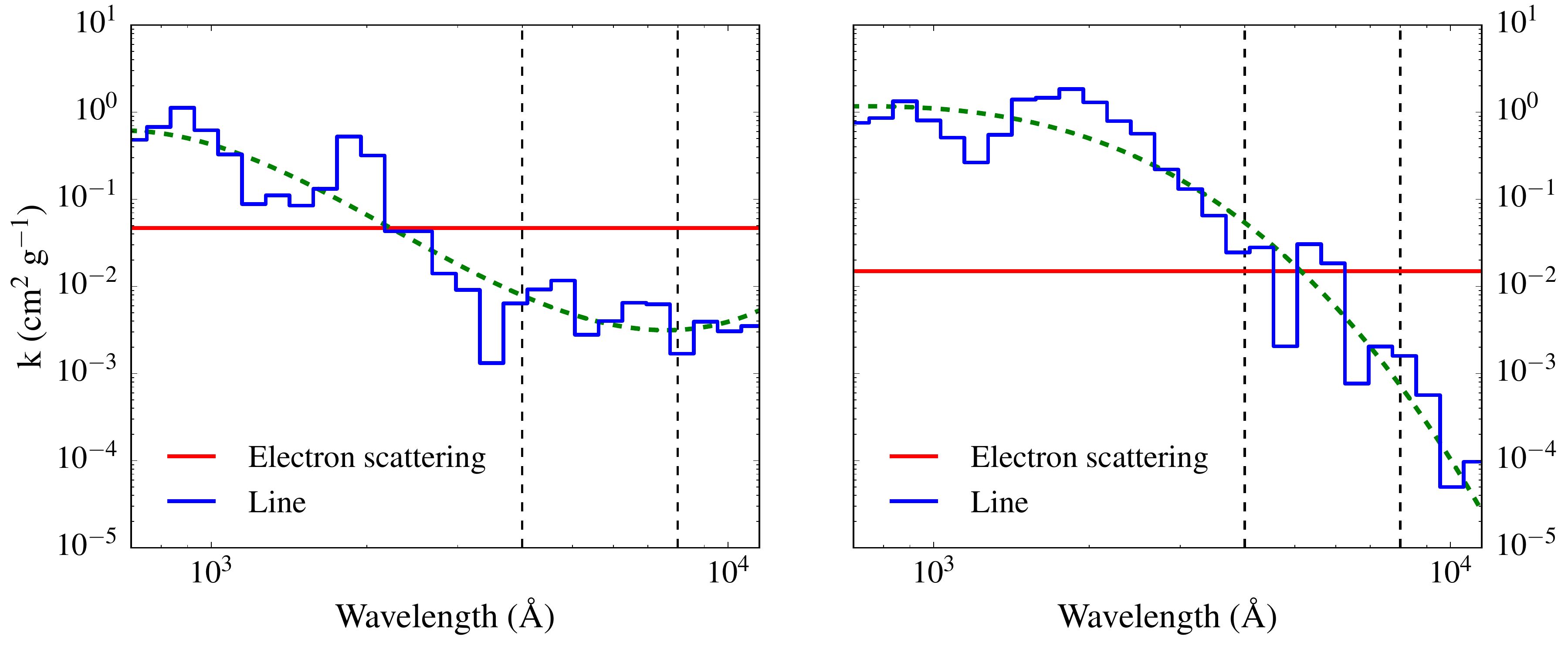}
\includegraphics[width=0.87\textwidth]{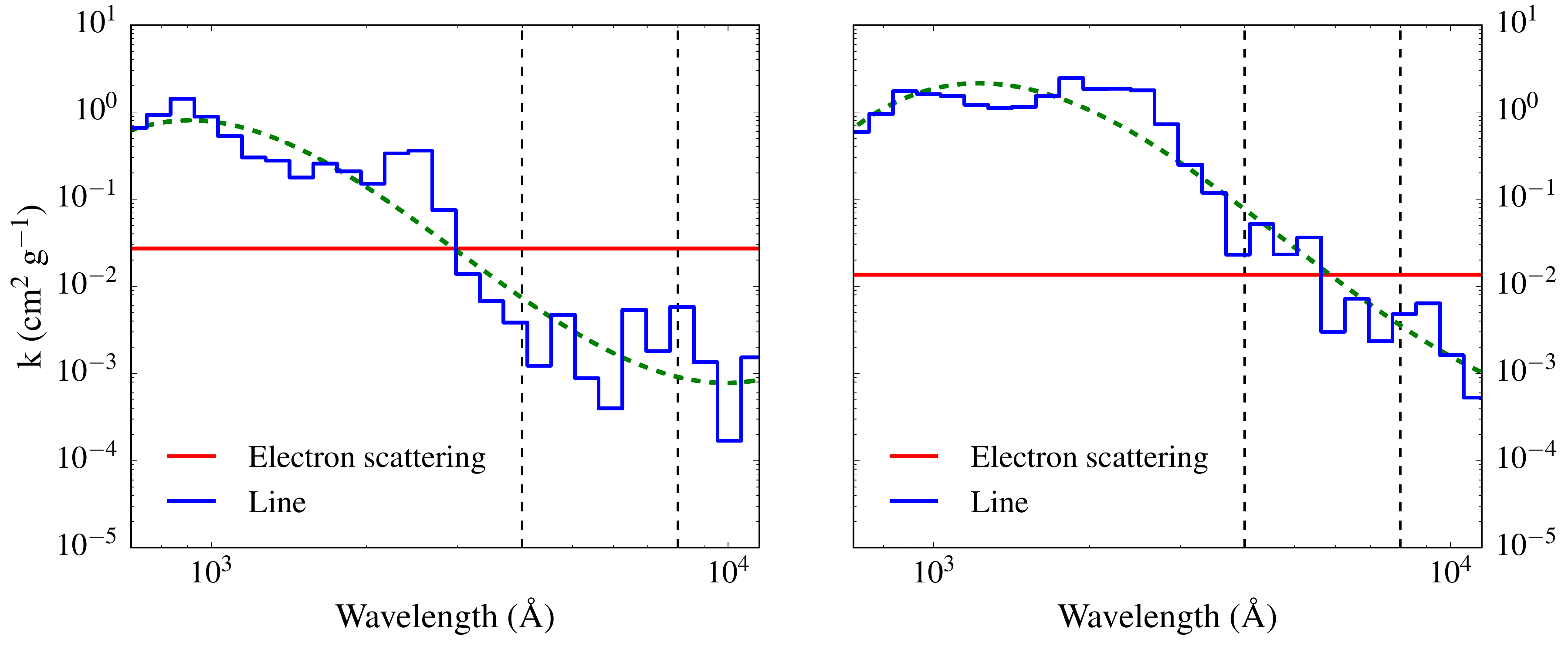}
\caption{Electron scattering (red) and line (blue) opacities extracted from the flux spectra using \textsc{tardis}. \textit{Left panels}: opacities extracted for the first (top) and second (bottom) epoch with an ejecta composition as suggested by \citet{maz16} (\textit{Model A}, Tab.~\ref{table:tardis}). \textit{Right panels}: same as the the left-hand panels but with the extreme case of pure Ni ejecta at the time of explosion (\textit{Model B}, Tab.~\ref{table:tardis}). A fit to the line opacity is shown as dashed green line and used in our polarization code to model a pseudo-continuum of line. Dashed black, vertical lines encompass the rest-frame wavelength region covered by our spectra.} 
\label{fig:tardis}
\end{center}
\end{figure*}

\begin{figure}
\begin{center}
\includegraphics[width=0.48\textwidth]{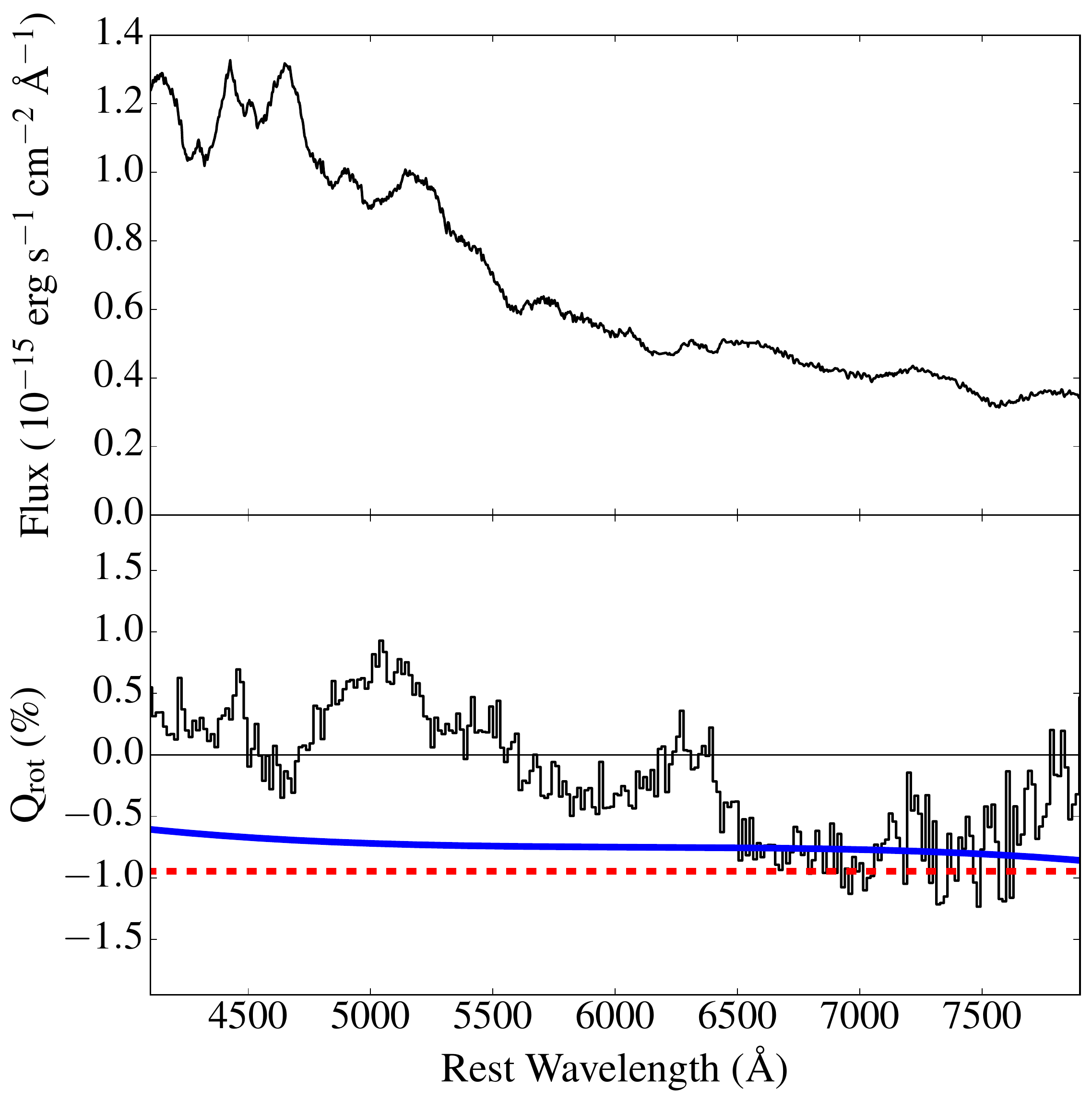}
\caption{Flux (top) and polarization (bottom) spectra of \ae\/ at 23.7~d before maximum light (black lines), together with polarization levels predicted by our toy model. A single ellipsoidal region with axis ratio \mbox{$A=0.88$} is used, while opacities are selected from \textit{Model A}. The red dashed line corresponds to a calculation including only electron scattering opacity, while the solid blue line shows to a calculation with both electron scattering and line opacities.} 
\label{fig:one_zone_fit}
\end{center}
\end{figure}

\subsection{Modeling}\label{sec:polmod}

\begin{figure*}
\begin{center}
\includegraphics[width=0.497\textwidth]{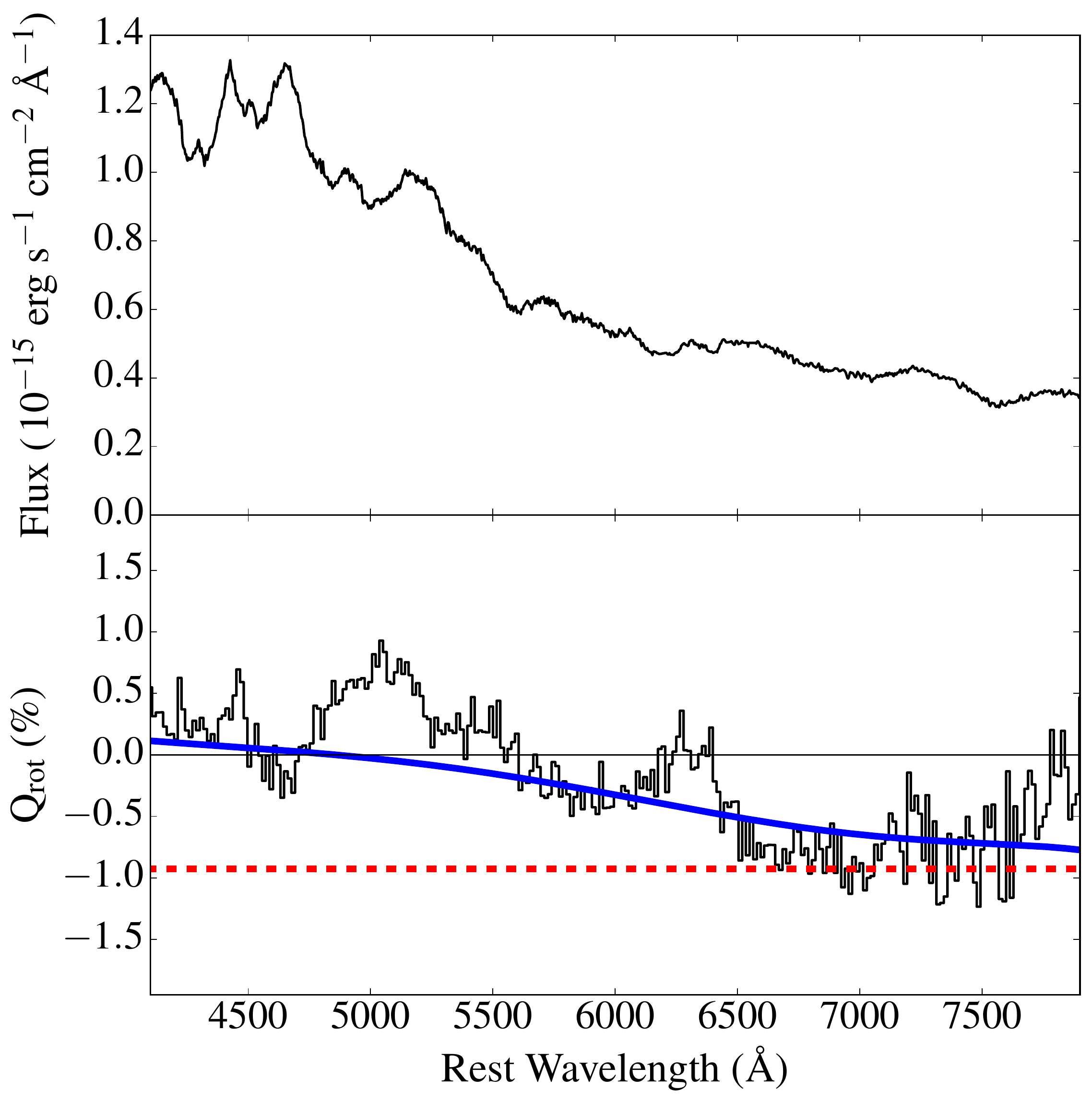}
\includegraphics[width=0.497\textwidth]{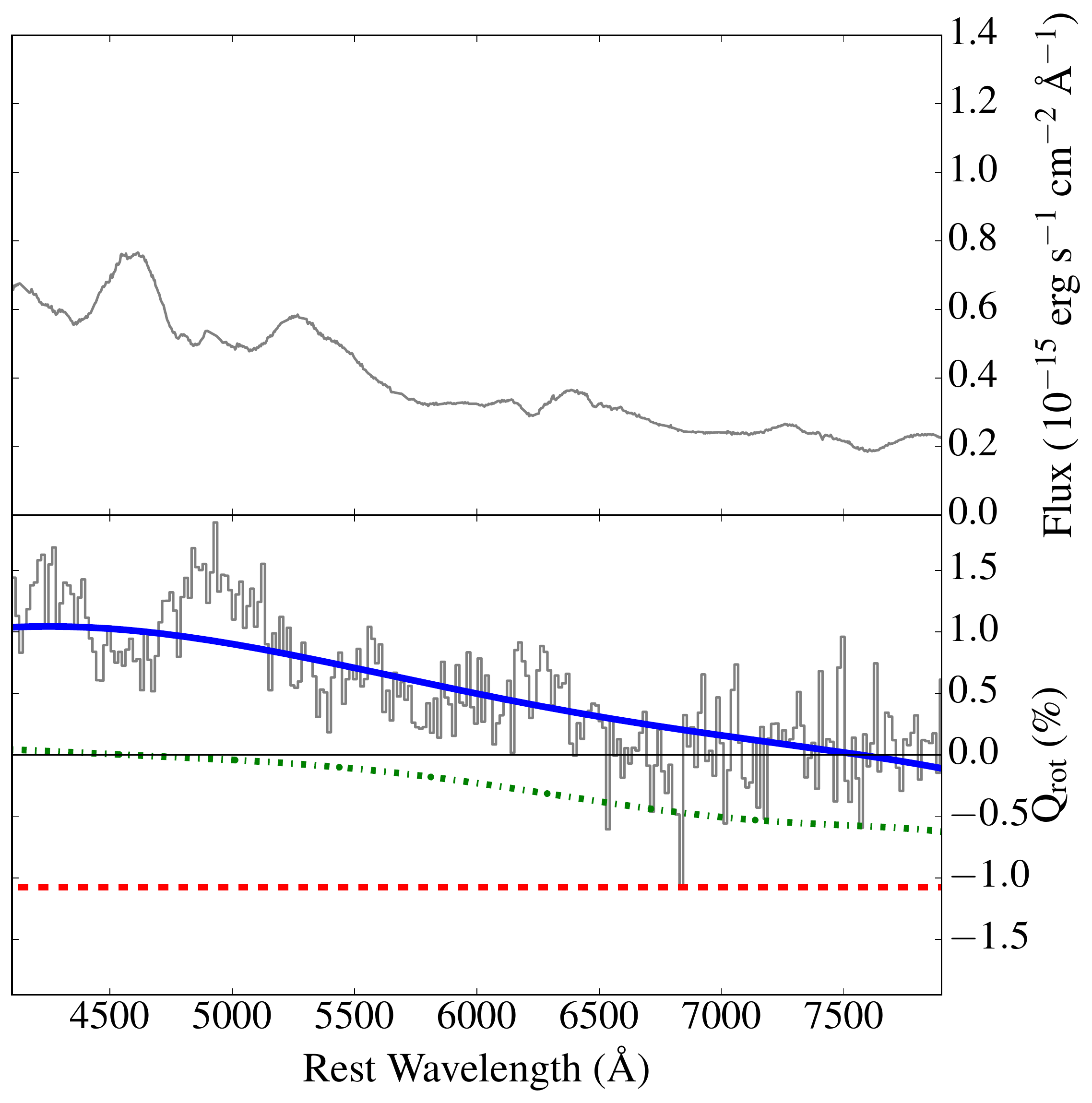}
\caption{\textit{Left panels:} flux and polarization spectra of \ae\/ at 23.7~d before maximum light (black lines), together with polarization levels predicted by our toy model. The red dashed line corresponds to a calculation including only electron scattering opacity, while the solid blue line shows to a calculation with both electron scattering and line opacities. The geometry used for both these calculations is discussed in the text and reported in the upper panel of Fig.~\ref{fig:geo}. \textit{Right panels:} flux and polarization spectra of \ae\/ at 27.5~d after maximum light (grey lines), together with different polarization predictions from our toy model.The red dashed line and green-dot dashed line are calculated adopting the same geometry used for the first epoch and excluding or including line opacities, respectively. The blue solid line assumes instead a different geometry (see lower panel of Fig.~\ref{fig:geo}) and includes both electron scattering and line opacities.} 
\label{fig:fit}
\end{center}
\end{figure*}

As discussed in the previous section, polarization spectra for \ae\/ are characterized by a well-defined axis of symmetry. In addition, the polarization component along the dominant axis, $Q_\textrm{rot}$, shows a strong wavelength dependence at both epochs and a pronounced evolution between pre- and post-maximum. In the following we present a simple toy-model that can account for such observed polarization signatures. We note that this model is only intended to characterise the observed data and that the results presented here may not be unique: similar results may be obtained with different geometries and compositions. Our primary goal is to
constrain the physical mechanisms responsible for (i) the wavelength dependence and (ii) the time evolution of the overall \textit{pseudo-continuum} polarization level, rather than to model polarization features associated with individual spectral transitions.

To interpret the polarization spectra for \ae\/ we will consider simple ellipsoidal geometries using the Monte Carlo toy-code presented by \citet{bulla2015}. Monte Carlo packets, representing bundles of photons, are created unpolarized at a spherical photosphere of radius $R_\textrm{ph}$, emitted assuming constant surface brightness and propagated throughout a prolate ellipsoidal envelope defined by
\begin{equation}
\frac{x^2}{a^2}+\frac{y^2}{b^2}+\frac{z^2}{c^2}=1~~,~~~~a=b<c~~~.
\end{equation}
The latter can be rewritten in cylindrical coordinates as
\begin{equation}
\frac{r^2}{A^2}+z^2=c^2
\end{equation}
where we have introduced the axis ratio $A=a/c<1$. Our calculations adopt ellipsoidal isodensity surfaces 
\begin{equation}
\rho(\xi,t)=\frac{\rho_0(t)}{\xi^n}~~,
\end{equation}
where
\begin{equation}
\xi=\sqrt{\frac{r^2}{A^2}+z^2}~~.
\end{equation}
For all our calculations we use a power-law index $n=7$ as suggested by previous studies \citep[e.g.][]{maz16}. On their journey throughout the ejecta, Monte Carlo packets  can interact with matter and change their polarization state. Two sources of opacity are used: a grey electron scattering opacity and a wavelength-dependent line opacity (see below). The former is assumed to partially polarize the radiation, while the latter regarded as a depolarizing contribution \citep[see e.g.][]{jeffery1989,jeffery1991}. Polarization spectra are extracted for an equatorial viewing angle (that is along the semi-minor axis of the prolate ellipsoid). Moving the observer orientation from the equatorial plane towards the pole would result in smaller polarization levels and thus would require larger asphericities than those predicted below.

The electron scattering coefficient and line opacity distribution adopted in our calculations are estimated 
using the 1D \textsc{tardis} radiative transfer code \citep{kerzendorf2014}. In particular, calculations with two different ejecta compositions have been carried out: one (hereafter referred to as \textit{Model~A}) adopting the ejecta composition proposed by \citet{maz16} and one (\textit{Model~B}) assuming an extreme case of $^{56}$Ni-dominated ejecta at the time of explosion. In the latter case, we derive the relative abundances of each element by calculating the fraction of Ni that has decayed to Fe and Co at both epochs of \ae\/. 
The specific ejecta compositions used for each calculation are reported in Table~\ref{table:tardis}.
\textsc{tardis} provides values for the electron scattering co-efficient and the Sobolev optical depth of each transition in each zone from which tables of wavelength-dependent expansion opacities can be calculated \citep{karp1977,friend1983,eastman1993}. 
The opacities obtained for \textsc{tardis} zones close to the photosphere in \textsc{tardis} are plotted in Fig.~\ref{fig:tardis}. At both epochs, \textit{Model~A} produces only a weak wavelength dependence in the line opacity distribution across the wavelength interval covered by our observations ($\sim$ 4000 to 8000~\AA), with values that are a factor of $\sim$10 lower than the electron scattering opacity ($k^{A,1}_\textrm{sc}=0.047~\textrm{cm}^2~\textrm{g}^{-1}$ and $k^{A,2}_\textrm{sc}=0.027~\textrm{cm}^2~\textrm{g}^{-1}$ at the first and second epoch, respectively). In contrast, the line opacity distribution extracted for \textit{Model~B} is strongly wavelength dependent in this region. While spectral regions around 8000~\AA{} are still dominated by electron scattering opacity ($k^{B,1}_\textrm{sc}=0.015~\textrm{cm}^2~\textrm{g}^{-1}$ and $k^{B,2}_\textrm{sc}=0.014~\textrm{cm}^2~\textrm{g}^{-1}$), the line opacity contribution steadily increases toward the bluer regions of the spectrum and becomes dominant around 4000~\AA.

Fig.~\ref{fig:one_zone_fit} reports results of two simulations relative to the first epoch of \ae\/. Both calculations adopt opacities from \textit{Model~A} and assume an ellipsoidal envelope with axis ratio $A=0.88$ and outer boundary at $\xi_\textrm{out}=2\times R_\textrm{ph}$. This value of the axis ratio was chosen to match the continuum polarization around 7000~\AA\/ (see below). The maximum electron scattering opacity to the boundary
\begin{equation}
\tau_\textrm{max}=\int_{R_\textrm{ph}}^{\xi_\textrm{out}}k_\textrm{sc}\,\rho(\xi,t)\,\textrm{d}z
\end{equation}
is set to unity. Ejecta are modeled 70~d after explosion and the photosphere placed at $v_\textrm{ph}^{1\textrm{st}}=8\times10^3$~\kms \citep{ni16}. The density at $\xi=R_\textrm{ph}$ and the mass in the envelope are $\rho^{1\textrm{st}}=2.7\times10^{-14}$~g~cm$^{-3}$ and $M^{1\textrm{st}}=2.5$~\M, respectively. The first polarization spectrum (red dashed line) corresponds to a calculation including only electron scattering ($k^{A,1}_\textrm{sc}$) and provides a good match to the polarization level detected for \ae\/ in the wavelength region around 7000~\AA{} that is devoid of strong optical features. The second polarization spectrum (solid blue line) shows instead to a simulation in which both electron scattering and line opacities are included. Specifically, we adopt a polynomial fit to the line opacity distribution as representative of the pseudo-continuum absorption component (see upper left panel of Fig.~\ref{fig:tardis}).
Compared to the line-free calculation, the latter produces lower polarization levels since the line opacity acts as a depolarizing contribution. This effect becomes stronger moving from redder to bluer wavelengths, as a consequence of the increasing contribution of line to the total opacity (see upper left panel of Fig.~\ref{fig:tardis}). The match between predicted and observed polarization levels, however, is poor as the wavelength dependence in the line opacity of \textit{Model~A} is not sufficiently strong to bring the overall $Q_\textrm{rot}$ level toward positive values below $\sim$~5500~\AA, as observed for \ae\/.  

Compared to \textit{Model~A}, line opacities extracted for \textit{Model~B} are characterized by a more pronounced wavelength dependence (see right panels of Fig.~\ref{fig:tardis}). Therefore, we construct a new model that includes opacities from both \textit{Model~A} and \textit{Model~B}. Specifically, we divide the envelope in two different ellispoidal zones: 
\begin{itemize}
\item A \textit{``metal-rich''} inner region modeled as a prolate ellipsoid with $\xi_\textrm{in}=1.5\times R_\textrm{ph}$. Opacities in this region are selected from \textit{Model B}. Specifically, we adopt $k^{B,1\textrm{st}}_\textrm{sc}$ (pre-maximum) and $k^{B,2\textrm{nd}}_\textrm{sc}$ (post-maximum) as electron scattering coefficients and use a polynomial fit to the line opacity distribution for the pseudo-continuum absorption component (see right panels of Fig.~\ref{fig:tardis}). 
\item A \textit{``metal-poor''} outer region modeled as a prolate ellipsoid with $\xi_\textrm{out}=2\times R_\textrm{ph}$. Opacities in this region are selected from \textit{Model A}. Here, we adopt electron scattering coefficient $k^{A,1\textrm{st}}_\textrm{sc}$ (pre-maximum) and $k^{A,2\textrm{nd}}_\textrm{sc}$ (post-maximum) and a polynomial fit to the line opacity distribution for the pseudo-continuum absorption component (see left panels of Fig.~\ref{fig:tardis}). 
\end{itemize}
The shapes of the two zones (i.e. the axis ratios $A_\textrm{in}$ and $A_\textrm{out}$) are taken as free parameters in our models, together with the maximum electron scattering opacity to the boundary $\tau_\textrm{max}$. A suitable choice for $\tau_\textrm{max}$ is selected (see below) and the axis ratios $A_\textrm{in}$ and $A_\textrm{out}$ chosen to reproduce the observed polarization spectra.

The left panel of Fig.~\ref{fig:fit} shows results of our simulations compared to the first epoch of \ae\/. As in the one-zone model presented above, these calculation are carried out at 70~d after explosion and assume $v_\textrm{ph}^{1\textrm{st}}=8\times10^3$~\kms and $\tau_\textrm{max}^{1\textrm{st}}=1$. The inner and the outer ellipsoidal envelopes have the same axis ratio, namely $A_\textrm{out}^{1\textrm{st}}=A_\textrm{in}^{1\textrm{st}}=0.88$. Estimated values for the density at $\xi=R_\textrm{ph}$ and for the mass in the envelope are $\rho^{1\textrm{st}}=7.3\times10^{-14}$~g~cm$^{-3}$ and $M^{1\textrm{st}}=6.7$~\M, respectively. The first simulation (dashed red line) includes only electron scattering in the ejecta and is found to reproduce the degree of polarization observed for \ae\/ around 7000~\AA ($Q_\textrm{rot}~\sim$~$-$1~per~cent).
While the grey electron scattering opacity assumed in the latter calculation yields a constant degree of polarization as a function of wavelength, including contributions from line absorption (solid blue line) results into a strong wavelength dependence of the pseudo-continuum level. This causes a sign reversal of $Q_\textrm{rot}$ around $5000$~\AA, in good agreement with what is observed for \ae\/. The similarity between our predictions and the overall pseudo-continuum level of \ae\/ thus suggests that an increasing contribution of line opacities towards the bluer regions of the spectra is responsible for the strong wavelength dependence observed.

The right panel of Fig.~\ref{fig:fit} shows results of our simulations relative to the second epoch of \ae\/. The latter is modeled 121~d after explosion and with the photosphere placed at $v_\textrm{ph}^{2\textrm{nd}}=7\times10^3$~\kms \citep{ni16}. Following the ejecta expansion, the density at $\xi=R_\textrm{ph}$ and the maximum electron scattering opacity drop to $\rho^{2\textrm{nd}}=3.6\times10^{-14}$~g~cm$^{-3}$ and $\tau_\textrm{max}^{2\textrm{nd}}=0.61$, respectively, while the mass in the envelope increases to $M^{2\textrm{nd}}=11.9$~\M. As shown in Fig.~\ref{fig:fit}, keeping the axis ratios of the ``metal-rich'' and ``metal-poor'' regions fixed to the values they have at the first epoch leads to polarization levels that are clearly inconsistent with those observed in \ae\/ (dashed-dot green line). If we instead keep the same axis ratio for the outer region ($A_\textrm{out}^{2\textrm{nd}}=A_\textrm{out}^{1\textrm{st}}=0.88$) and assume a more aspherical inner region ($A_\textrm{in}^{2\textrm{nd}}=0.60$), we obtain a polarization spectrum in good agreement with data (solid blue line). 

\begin{figure}
\begin{center}
\includegraphics[width=0.5\columnwidth]{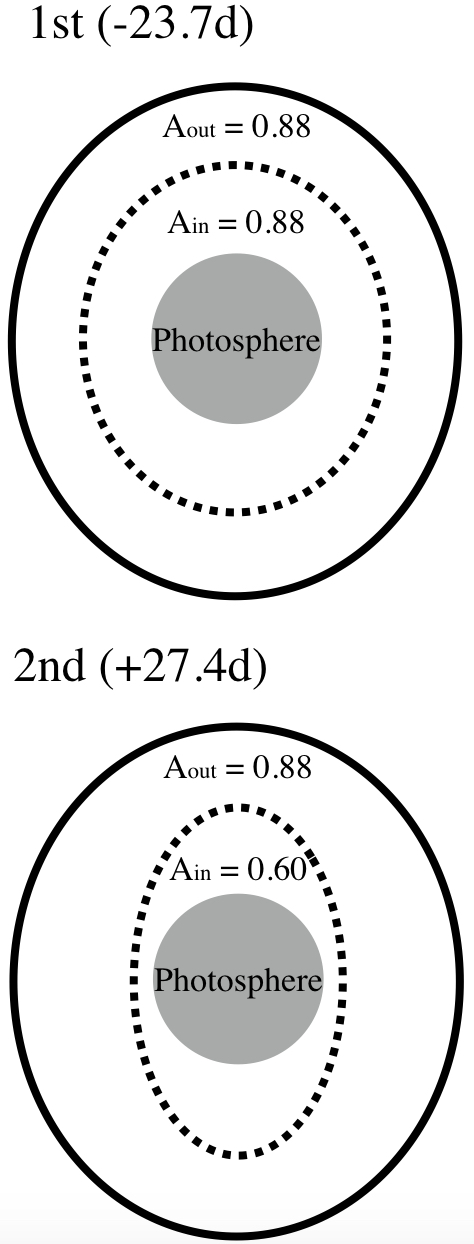}
\caption{Possible geometry of the ejecta at the first, pre-peak epoch and at the second, post-peak epoch. The external layer retains its prolate geometry (A$_{\rm out}= {\rm a/c} = 0.88$), while the inner shell has increased its asphericity from A$_{\rm in}=0.88$ to A$_{\rm in}=0.60$. The pictured ellipsoids have the same asphericity reported in the figure and in the text.} 
\label{fig:geo}
\end{center}
\end{figure}

To summarize, our ellipsoidal toy model suggests that (i) the strong wavelength dependence observed in both polarization spectra of \ae\/ is given by a strong increase in the line opacity from redder to bluer regions of the spectra, which has been previously predicted for models of type II SNe \citep{de11}; (ii) the time evolution of the polarization level from pre- to post-maximum is given by a change in the asphericity of the inner layers (from $A_\textrm{in}^{1\textrm{st}}=0.88$ to $A_\textrm{in}^{2\textrm{nd}}=0.60$). The ejecta geometries proposed by our calculations are reported in Fig.~\ref{fig:geo}. The implications of these results for the different progenitor scenarios will be discussed in Section~\ref{sec:dis}. 

We stress again that the simple ellipsoidal modeling developed here is only intended to broadly characterize the data and frame their interpretation. In particular, we did not attempt to model any discrete spectral features nor investigate the extent to which the two epochs are consistent with evolution in homologous expansion. Instead we have considered the two epochs individually.

\section{On the geometry of superluminous supernovae}\label{sec:dis}
Several scenarios have been proposed to account for the 
exceptional luminosity of SLSNe. Among these, the most successful to reproduce light curves, as well as velocity and temperature evolution, has been that of a central engine, likely a magnetar, that deposits its energy into the supernova explosion \citep{kb10,wo10,in13,ni13}. However, other scenarios including the interaction of the ejecta with CSM due to a pulsational pair instability SN \citep[PPISNe, e.g.][]{woo07} or a black hole as central engine \citep{dk13} cannot be ruled out to explain the variety of SLSNe observables.
Our spectropolarimetric observations provide a new insight into the geometry of a SLSN explosion and we can now consider the implications of this for identifying the explosion mechanism responsible for their extreme luminosity. 

The data show a deviation from spherical symmetry and the presence of a dominant axis which implies an axisymmetric configuration of the ejecta.
Furthermore, the U$_{\rm rot}$ polarization vector in the new rotated system - where the Q$_{\rm rot}$ vector is along the dominant axis and the U$_{\rm rot}$ is orthogonal to the previous (see Sect.~\ref{sec:rot}) - indicates that there is almost no physical deviation from the dominant axis. 
To explain the observed increase in the Q$_{\rm rot}$ vector from the pre-peak to the post-peak epoch (see Fig.~\ref{fig:rot})  we need  an increase in the asphericity of the inner layers of the ejecta or the photosphere to recede into inner layers 
which are less spherical.
The geometric parameters inferred from our toy ellipsoidal modeling 
of the inner zone (A$_{\rm in}^{1}=0.88$ and A$_{\rm in}^{2}=0.60$, see Sec.~\ref{sec:polmod}) are certainly model dependent and so can only be taken as indicative: they depend in detail on the composition/line opacity adopted and on all parameters controlling the shape and density distribution. To fully understand degeneracies among these parameters, and to construct a self-consistent model that accounts for both epochs is beyond the scope of the simplistic approach used here. Nevertheless, the calculations illustrate that moderate departures from sphericity combined with substantial wavelength-dependent opacity can qualitatively account for the observed spectropolarimetric properties.
Although we used ellipsoidal toy models to characterize the data (shown in Fig.~\ref{fig:geo}), it is important to note that alternative axi-symmetric  
configurations, including bipolar ejecta morphologies, 
would also be able to match the observations.

Together, the data and modeling presented above  suggest that, as time passes, the spectral forming region recedes into increasingly aspherical (but axi-symmetric) ejecta layers. In the following, we consider the ramifications of this conclusions for our understanding of SLSN explosion models, bearing in mind models involving a dominant axis are  favored by the data. 

Multidimensional simulations of magneto-rotational core-collapse SNe have already shown large scale asymmetries \citep{mo14}.
In the case of a magnetar as the inner engine, the overall axisymmetry and the increase of asphericity as the photosphere recedes through the inner layers with time could be the consequence of several factors.

\begin{enumerate}
\item  It could be a consequence of a jet-like flow stalled within the core in a similar fashion to that proposed for SN~2008D/XRF~080109 \citep{ma09}. This is an event linked to  X-ray flashes which may 
also be related  to $\gamma$-ray bursts but with a softer radiation outburst. Hypothetically, this suggestion could also be supported by the possible link between SLSNe Ic and ultra long $\gamma$-ray burst \citep{gr15,me15,ka16}. 
\item  As the ejecta expands, its density decreases and the non-thermal radiation from the magnetar wind nebula can 
ionize more of the ejecta \citep{me14}. A magnetar wind nebula that is asymmetric could enhance such behavior. 
\item 2D simulations of magnetar powered SLSNe \citep{chk16} have shown that fluid instabilities and mixing occurs
if energy  is deposited in a relatively small mass of material which could be,  approximately, the mass of the inner ejecta. 
The amount of energy deposited needs to be comparable to that of the ejecta kinetic energy of the explosion. 
These instabilities and mixing could alter the geometry of the ejecta, giving an axysimmetric geometry reminiscent of 
 that of the Crab nebula, if the initial magnetar period is less than 3 ms \citep{chk16}.
Notably, the magnetar models proposed for the bolometric light curve of \ae\/ do satisfy this condition \citep[2.1 ms and 1.7 ms for the two best fit shown in][]{ni16}.
\end{enumerate}

In the case of a black hole as a central engine, the first scenario described above (the jet like flow) would still be applicable. 
To date, there have been no multidimensional simulation of this mechanism 
and little can be added about the geometry of the ejecta. However, the inner engine mechanism would need to explain the wavelength dependence observed in \ae\/ data. 

Recently \citet{koz15} have suggested that pair instability SNe (PISNe) could still be a viable scenario to explain slow evolving SLSNe Ic like \ae\/.  Ejecta from such explosions have a metal rich inner ejecta. 
The attempt in modeling the wavelength dependence in our spectropolarimetry data suggests a strong line opacity, achievable with a very metal rich composition. However, there is some tension between this ejecta composition and the observed spectra since they do not show very obvious signatures of strong ling blanketing. This suggest as detailed modelling including self-consisting spectra and polarisation is needed. 
Additionally, it is hard for such massive explosions to break the original spherical symmetry as shown by several multidimensional simulations \citep[][]{fr01,jw11,ba13,ch14pisne}.
 A multidimensional simulation including rotation was actually able to produce an axisymmetric inner ejecta for a highly rotating CO core \citep{cha13}, but this was at the cost of lower luminosity and redder spectra than those of non-rotating PISNe.  We note that one 
object showing similar behavior to that of highly rotating PISNe has been reported \citep[PS1-14bj,][]{lu16}.
However, a PISN explosion for \ae\/  is hard to reconcile with all the data we have to date \citep[see also][]{je16}.

The last scenario to consider is that of the CSM interaction. As mentioned in Section~\ref{sec:intro}, the most promising mechanism to explain the high luminosity in the framework of interaction is that of a pulsational pair instability SN (PPISN), where thermonuclear outbursts due to a recurring pair-instability release massive shells of material that eventually collide with each other. The undulations observed in the \ae\/ light curve\citep{ni16}, as in other slow-evolving SLSNe Ic \citep[][Inserra et al., in preparation]{lu16},  could be a consequence of collisions between multiple shells.
Such collisions could lead to a strong mixing and Rayleigh-Taylor instabilities creating a structure which deviates from spherical symmetry behind the shock generated in the last collision \citep{ch14ppisne}. However, such instabilities would not preserve an axis. Additionally, the outermost collision is like that between a SN ejecta and a dense shell expelled by the star prior to its death as in the case of type IIn. This was suggested for the case of SLSN IIn SN~2006gy \citep{woo07}, which shows narrow H lines in the spectra throughout its evolution. 
Hence, it is difficult to reconcile the flux and polarization spectra with those of interacting SNe. Indeed, even if we assume that in our H- and He-free environment we should not observe narrow lines belonging to the unshocked external material. 
We should have observed an overall polarized continuum as in the other interacting supernovae \citep[e.g.][]{pa11} and a line polarization similar to those observed in normal, non-interacting SNe which show an axisymmetric photosphere below the interaction.
As reported in Sections~\ref{sec:sp}~\&~\ref{sec:spol} and shown in Fig.~\ref{fig:original} we do not observe such behavior.   

\section{Conclusions}
We have presented the first spectropolarimetric data for a SLSN. We have gathered two epochs, one pre-peak at -23.7d and the second 27.5 days after maximum light in the rest-frame. 

Our analysis of these data indicates:
\begin{enumerate}
\item the presence of a dominant axis and no physical departure from it;
\item a strong wavelength dependence of the polarization spectra; and
\item an increase in the mean degree of polarization from the first to the second epoch of observations
\end{enumerate}

We used our Monte Carlo toy-code to compare with the data in order to interpret 
 the wavelength dependence and the time evolution of the polarization level.  We calculated polarization spectra for prolate ellipsoidal geometries accounting for electron scattering and resonant line scattering. We were able to reproduce the wavelength dependence of the polarization by adopting a line opacity distribution  for the inner ejecta that is rich in iron-group elements. This poses some challenges for the observed flux spectra, but further comprehensive modelling 
with full radiative transfer is required.
We have also show that the evolution of the overall pseudo-continuum level can be replicated via two-zone aspherical ejecta models with an outer zone having A$_{\rm out}=0.88$ at both epochs and an inner zone showing an increase in the asymmetry from the first (A$_{\rm in}=0.88$) to the second (A$_{\rm in}=0.60$) epoch. These values are calculated for an equatorial viewing angle and thus should be considered as lower limits on the asphericities of the ejecta. Orientations away from the equatorial plane would lead to lower polarization signals and would then require smaller axis ratio values. In addition, we caution that the specific values derived here (A$_{\rm in}=0.88-0.60$ and A$_{\rm out}=0.88$) are model dependent. However, our findings qualitatively demonstrate that the \ae\/ spectropolarimetry can be reproduced by an ellipsoidal geometry - or alternatively a bipolar geometry - with an inner region that increases its asymmetry as time passes.

The general trend for an increase in polarization with time (and the implied geometry) are reminiscent of those observed in stripped-envelope core-collapse SNe connected with $\gamma$-ray bursts, supporting the possible link between SLSNe Ic and ultra long $\gamma$-ray bursts. Among all the suggested scenarios for SLSNe I, these new data tend to favor a core-collapse explosion and a central inner engine as explanation for the inferred axisymmetric geometry, as well as the increase of the inner asymmetry. This could be achieved with an explosion of a rotating stellar progenitor. Currently, this scenario does not comfortably reproduce the wavelength dependence we observe, since it seems challenging to obtain enough line opacity. On the other hand, we did not explore other opacity options than a simple iron core and hence a more complete modelling is needed (including self-consistent spectra and polarimetry).

Despite the uncertainties that still exist with our relatively simple modelling approach,  the data 
should drive new initiatives in how to model these data with hydrodynamic simulations. These models will 
address questions such as the the geometry (bipolar or ellipsoidal shapes) and 
how an engine driven or interaction scenario could explain the wavelength dependence we observe 
(which we interpret as due to line opacity). We also need to know if
 other SLSNe I and broad line SLSNe II show similar properties. 
The next  step is to collect further observations of the future nearby ($z<0.2$) SLSNe at least at two epochs, ideally one before and one post maximum light. Predicting polarization signatures for multi-dimensional hydrodynamic explosion models will 
then be  required to better investigate the geometry of these explosions. 

\acknowledgments{
This paper is based on observations made with ESO Telescopes at the Paranal Observatory under program ID 095.D-0611(A).
CI thanks the ESO staff of the VLT, and especially Joe Anderson, for his competent support of this project in service mode and to have carried out these great observations. CI also thanks Matt Nicholl to have kindly provided information about \ae\/ before the submission of his paper and for comments on SN~2015bn. CI and MB thank Emma Reilly, Ferdinando Patat and Stefan Taubenberger for useful discussions on the reduction and errors evaluation of spectropolarimetric data. CI also thanks Ting-Wan Chen and Anders Jerkstrand for useful insights on the topic of SLSNe.
We thank PESSTO, (the Public ESO Spectroscopic Survey for Transient Objects Survey) ESO programs 188.D-3003,  191.D-0935 for the prompt classification of the object. 
SJS acknowledges funding from the European Research Council under the European Union's Seventh Framework Programme (FP7/2007-2013)/ERC Grant agreement n$^{\rm o}$ [291222] and  STFC grants ST/I001123/1 and ST/L000709/1. 
}

\bibliographystyle{yahapj}

\end{document}